\documentclass[10pt] {IEEEtran}
\usepackage{tikz}

\usepackage{graphicx,array,amsmath}
\usepackage{subfig,url}
\usetikzlibrary{positioning,shapes}
\DeclareMathOperator\erfc{erfc}

\begin{document}

\title {Empirical model for combinatorial data center network switch design}  

\author{\textsf{Ganesh C. Sankaran$^{1}$, Pachava Srinivas$^2$, Balaji
    Srinivasan$^2$ and Krishna M. Sivalingam$^3$} 

\footnotesize 
$^1$HCL Technologies Ltd, Chennai, INDIA\\
$^2$Department of Electrical 
    Engineering, Indian Institute of Technology Madras, Chennai,
    INDIA\\
$^3$Department of Computer Science and
    Engineering, Indian Institute of Technology Madras, Chennai,
    INDIA\\
Email: \textit{gsankara@hcl.com, balajis@ee.iitm.ac.in,
  skrishnam@iitm.ac.in, krishna.sivalingam@gmail.com}
	}
\maketitle

\begin{abstract}
Data centers require high-performance network equipment that consume
low power and support high bandwidth requirements. In this context, a
combinatorial approach was proposed to design data center network
(DCN) equipment from a library of components in \cite{infocom}. This
library includes power splitter, wavelength multiplexers,
reconfigurable add-drop multiplexers and optical amplifiers. When
interconnecting optical components, it must be ensured that the
resultant network supports specified target bit-error-rates
(typically, at most $10^{-12}$). This paper reports experiment
conducted on component interconnections and their computed
bit-error-rates. From the experimental analysis, it was observed that
the desired objective can be decided by considering a zeroth-order
threshold for optical power at the receiver and before the
amplifier. This paves way for the theoretical evaluation of several
other such designs using this empirically derived model.
\end{abstract}

\section{Introduction}

Data centers centralize compute and storage requirements of an
enterprise or a service provider.  A task in a data center is
typically performed in a distributed manner using a set of
compute-and-storage (CSN) nodes. The interconnection network, called
the data center network (DCN), that connects these nodes has a
significant impact on task completion times. Further, low power
consumption and high scalability (in terms of number of CSN nodes) are
crucial system design requirements.

Several network architectures have been proposed \cite{lion, eodcn,
  ogdcn} to satisfy high-performance, high-scalability and low power
objectives. A formal approach to design a high-performance and low
power data center network was proposed in \cite{infocom}. This was
modelled as a constraint optimization problem (CoP). This used a
combinatorial approach and explored all possible component sequences
to identify the best possible sequence. It involved evaluation of
several thousand component sequences for feasibility. It is not
feasible to study the individual sequences using detailed simulations
or using experiments. Thus, owing to large volume of inputs, the
evaluation must be largely theoretical.

The combinatorial solver uses a library of components to create the
component sequences \cite{solver}. These components include power
splitters, combiners, wavelength multiplexers, demultiplexers,
reconfigurable add-drop multiplexers, wavelength routers, optical
amplifiers and transceivers. When dynamically constructing a component
sequence from these components, it must meet the stringent optical
domain requirements.

One of the critical decisions to be made by the combinatorial solver
is to decide whether the chosen network can operate with tolerable bit
error rates (BER). This decision is known as BER satisfiability
decision (BSD). BSD must be made for all networks that are be created
by combining a set of components. A typical data center network is a
(relatively) short distance multi-fiber network. Though theoretical
models are available for long distance single-fiber spans with
amplifiers, the specific class of short distance multi-fiber networks
are not experimented widely to the best of our knowledge. Thus,
associated theoretical models are not readily available.

This paper attempts to address this gap by conducting experiments and
thereby derives an empirical model. A small set of networks are
experimentally created. Optical power is measured for these networks
at all points and the received signal is recorded. The BER is computed
by analyzing this recorded signal. The decision tree algorithm, which
is popular in analytics, is used to arrive at BSD with optical power
levels as its input. Networks can be designed using passive optical
components alone or using passive components along with an
amplifier. Both these network designs are experimentally studied in
this paper.

From the analysis it is observed that the optical power at the
receiver and before the amplifier influence the BSD
decision. Interestingly, these factors are also part of the
single-fiber long distance model. Finally, BSD can be made by
considering two optical power level thresholds.

\section{Solver description}

Current data center networks must satisfy many requirements
simultaneously. These requirements include power consumption,
throughput and latency. Researchers have proposed data center networks
that satisfy one requirement at a time. If a proposal outperforms
others on latency, it is often outperformed on another
requirement. Thus, it is difficult to build a one-size-fits-all
network that satisfies a wide range of requirements. Hence,
custom-designed networks that satisfy the given set of requirements
are needed.

When these requirements are encoded as constraints, a custom-built
constraint optimization problem (CoP) solver \cite{solver} finds the
best possible solution. This solver explores a large N-dimensional
search space for solutions that satisfy the constraints. Then, based
on the objective the best solution is identified. A solution
identified by the solver is a component sequence. This sequence is
built from a library of discrete optical components and their
characteristics. The solver attempts to find the optimal sequence made
up of these optical components. Many component sequences must be
explored to find the best possible component
sequence. Being combinatorial, several thousand component sequences
must be explored to find the best possible component
sequence for large networks.

Evaluating these component sequences is a tough task. There are three
approaches widely used for evaluation namely: theoretical, simulation
based and experimental evaluation. Solver adopts theoretical
evaluation. This approach is well suited for exploring
large search spaces in minimum time. However, a suitable theoretical
model must be available to the CoP solver \cite{solver} to take
critical decisions. 

%All optical components of required number must be available for experimental evaluation of these sequences. Once these components are available, thousands of component sequences must be created. Time required for creating all these component sequences is prohibitively high. Simulation based evaluation is suitable for small number of networks but this is also not suitable for evaluating thousands of component sequences. 

The DCN is a relatively short distance network that typically spans a few Km
at most. Thus, the corresponding component sequence is also a
short-distance network. Optical domain characteristics
must be modelled for this network and this
must be encoded as constraints. This ensures that the optimal sequence
identified satisfies critical optical domain constraints such as
bit-error-rate (BER) requirements. The library of components contains 
one-to-many (e.g. splitters) and many-to-one (e.g. combiners). These components
work with multiple fibers. While, long distance single-fiber networks are
studied widely including in \cite{ashwin}, a similar model for
short-distance multi-fiber network is not readily available. This
paper explores this aspect and attempts to model this specific class of networks.

\section{Experiment description}

The objective of the experiment is to empirically decide on whether a
network passes the BSD test or not. In other words, networks that have
a BER of less than $10^{-12}$ are accepted and the rest are rejected
by the BSD test. A set of experimental networks were created using
Lightrunner kit \cite{lightrunner} and their BER was computed. The
signal was modulated with data at 2.5~MHz.

The power level before and after every component was measured. This
was measured in addition to measuring the power levels at the transmitter
and the receiver. The difference in power levels before and after a
passive component is equal to the loss inserted by the component. For
an amplifier, the difference in power levels is the amplifier gain.  

Different components were used to create a component sequence. It 
includes a $1 \times 2$ power
splitter, a $2 \times 1$ power combiner, a wavelength multiplexer, 
a wavelength demultiplexer and an Erbium doped fiber amplifier (EDFA)
amplifier.

\subsection{BER estimation}

The signal received for every network is recorded in persistence
mode. This signal is then fed to MATLAB curve fitting tool to fit the
raw signal data to a double Gaussian curve. Let $\mu_1$ and $\mu_0$ be
the estimated mean for bit one and bit zero. Similarly, let $\sigma_1$
and $\sigma_0$  be the estimated standard deviation for bit one and
bit zero. The Q factor of the signal is given by $\displaystyle Q =
\frac{\mu_1 - \mu_0}{\sigma_1 + \sigma_0}$. Then, the corresponding
BER is given by $\displaystyle BER = \frac {1} {2} \erfc \left(
{\frac{Q}{\sqrt{2}}} \right) $. 

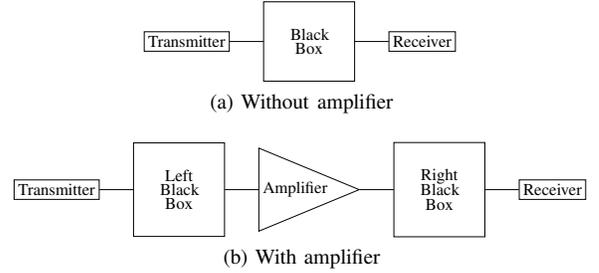
\begin{figure}[htbp]%
\centering
\subfloat[Without amplifier] {\scalebox{0.45}{\begin{tikzpicture}
\node[draw, align=center] (tx) {\Large Transmitter};
\node[draw, inner sep=5ex,text width=7ex,align=center,right=of tx] (lbb) {\Large Black Box};
%\node[draw, isosceles triangle,align=center,right=of lbb] (amp) {Amplifier};
%\node[draw, inner sep=5ex,text width=7ex,align=center,right=of amp] (rbb) {Right Black Box};
\node[draw, align=center,right=of lbb] (rx) {\Large Receiver};
\draw (tx) -- (lbb) -- (rx);
\end{tikzpicture}}}

\subfloat[With amplifier] {\scalebox{0.45}{\begin{tikzpicture}
\node[draw, align=center] (tx) {\Large Transmitter};
\node[draw, inner sep=5ex,text width=7ex,align=center,right=of tx] (lbb) {\Large Left Black Box};
\node[draw, isosceles triangle,align=center,right=of lbb] (amp) {\Large Amplifier};
\node[draw, inner sep=5ex,text width=7ex,align=center,right=of amp] (rbb) {\Large Right Black Box};
\node[draw, align=center,right=of rbb] (rx) {\Large Receiver};
\draw (tx) -- (lbb) -- (amp) -- (rbb) -- (rx);
\end{tikzpicture}}}

\caption{Schematic diagram for the experiments.}%
\label{fig:schematic}%
\end{figure}

\section{Experimental Data Analytics}
%% Based on this BER metric, the experimental networks are classified as
%% \textit{good}, corresponding to the ones that have a BER less than
%% $10^{-12}$ and \textit{bad}, otherwise.  

A decision tree algorithm,
available in the R \cite{R} package, was used for data analytics. The
inputs for this algorithm are the measured values and the output is
the classification of the network based on its BER.

Using optical components, two types of DCN designs are possible. The
first one uses passive optical components between a pair of
transceivers, as shown in Fig.~\ref{fig:schematic}(a). The second one
additionally also an amplifier Fig.~\ref{fig:schematic}(b). Both these
scenarios are experimentally studied. All experimental networks have a
transmitter and receiver on either sides. One or more black-boxes are
presented in the schematic diagram. During the study, these
black-boxes are replaced by actual optical components.

\subsection{Without amplifier}

A set of scenarios were created without an amplifier between the
transmitter and the receiver. The corresponding schematic is presented
in Fig.~\ref{fig:schematic}(a). In this case, two wavelengths (1510
and 1550~nm) were launched and measured separately. Signal power
was recorded at all points as described before. The received signal 
was recorded. This was used to estimate the corresponding BER.

\begin{table}
\centering
{\renewcommand{\arraystretch}{1.1}
\begin{tabular} {|l|l|l|r|r|}
\hline
S.No & BB & $\lambda$ in nm & Rx Power (in dBm) &BER
\tabularnewline \hline 
1 & SS & 1510 &  \textbf{-16.25} & 4.08E-21 \tabularnewline \cline{3-5}
 &  & 1550 &  -16.65 & 1.9E-08 \tabularnewline \hline 
2 & MM & 1510 &  \textbf{-11.45} & 1.22E-21 \tabularnewline \cline{3-5}
 & & 1550 &  \textbf{-10.55} & 2.24E-15 \tabularnewline \hline 
3 & SMMS& 1510 &  -17.15 & 2.43E-06 \tabularnewline \cline{3-5}
 & & 1550 &  -18.95 & 3.57E-09 \tabularnewline \hline 
4 & SMSSMS &1510 &  -28.15 & 4.20E-04 \tabularnewline \cline{3-5}
 & &1550 &  -25.65 & 1.48E-05 \tabularnewline \hline 
5 & SMSMMSMS & 1510 &  -29.65 & 4.23E-03 \tabularnewline \cline{3-5}
 & & 1550 &  -32.25 & 1.81E-03 \tabularnewline \hline 
\end{tabular}
}
\caption{Experiment scenarios without amplifier: M denotes wavelength
  multiplexer or demultiplexer and S denotes power combiner or
  splitter. The sequence of letters indicates an interconnection of
  components in the same order. Rx power is the received power.} 
\label{tbl:edat2}
\end{table}

Table~\ref{tbl:edat2} presents the scenarios that were 
experimented. The contents of the black box in Fig.~\ref{fig:schematic}
(a) is presented in the second column. 
Subsequent columns present the transmission wavelength 
used for the experiment, received power measured and 
estimated BER values respectively. Every row presents an 
experimental scenario. Though many other power values 
were recorded, they are not shown here for brevity. For 
instance, the black box of the first row contains \textit{
 SS}. It has two optical splitter or combiners connected 
back-to-back. When 1510~nm is transmitted from the left 
side of the black-box,  the received power on the right 
side is observed to be -16.25~dBm. The signal received is 
recorded and BER is estimated. The corresponding BER is $4
 .08 \times 10^{-21}$. This error rate is less than the 
tolerable BER. Considering third row of the table, the 
black box contains an optical combiner, a wavelength 
multiplexer, a wavelength demultiplexer and then an 
optical splitter from the left to the right side. When 
1510~nm is launched at the transmitter, the received 
power is observed to be -17.15~dBm and the corresponding 
estimated BER is $2.43 \times 10^{-6}$. The error rate is 
more than the tolerable BER and hence, given its received 
power, this black box is not a valid combination. 
Similarly, other entries in the table are examined. 
Examining all the entries in the table, it is observed 
that when the received power is at least -16.25~dBm, the 
corresponding BER is less than tolerable BER.

%% Ganesh - state that is for SS model, correct? Pl discuss one other
%% row, say row 4. Please check if the highlighted values are correct.
%% Pl. do the same for Table II. 

%% We can cut down stuff from Section I. 
%% Also, the short-distance link is described many times.

\begin{figure}%
\centering
{\scalebox{0.8}{\usetikzlibrary{positioning}
\begin{tikzpicture}
\node[ellipse,text width=16 ex,draw,align=center] (p) {Power before amplification \\ p=0.015};
\node[draw,rectangle,below=of p] (n2) at (-2.5,-0.6) {Node 2 (n=7)};
\node[draw,rectangle,below=of p] (n3) at (2.5,-0.6) {Node 3 (n=16)};
\draw (p) -- (n2.north west) node[midway,fill=white] {$\leq 26.38$};
\draw (p) -- (n3.north east) node[midway,fill=white] {$> 26.38$};
\end{tikzpicture}}}
\caption{Empirical model as decision tree algorithm's output.}%
\label{fig:emp}%
\end{figure}
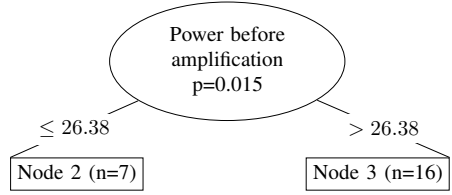

\subsection{With amplifier}
These networks have an amplifier sandwiched between two
black-boxes. The other end of the left and right black-boxes is
connected to a transmitter and a receiver respectively as shown in
Fig.~\ref{fig:schematic}(b). 

In the experimental network, the black-boxes are replaced with one or
more passive optical components. The left black box cannot be empty
but the right black-box can be empty. When this black-box is empty,
the output of the amplifier is directly connected to the receiver.  

%Initially logistic regression is used to identify and eliminate the set of unrelated inputs. For instance, the individual power levels did not have any correlation with the classification output. 

Initially power levels observed at all stages and the corresponding BER were subjected
to visual correlation. Then a subset of power levels that had good
correlation were fed as input to the decision-tree algorithm. This
decision tree is shown in Fig.~\ref{fig:emp}. From this decision tree
it can be observed that BSD can be made by considering whether power
level before amplification is less than -26.38~dBm. This empirical
model or the decision tree can be directly embedded into the
combinatorial solver to evaluate other component interconnections.  

Data pertaining to different experimental scenarios is presented in
Table~\ref{tbl:edat}. The table presents the power level before
amplification and the calculated BER for every scenario along with the
composition of its left and right black boxes. An empty right black
box can be seen in some scenarios. This indicates that the amplifier's 
output was connected directly to the receiver's input using a fiber.

Let us consider row 4 from Table~\ref{tbl:edat}. In this scenario, two wavelength demultiplexers (denoted by MM)were connected between the transmitter and the amplifier's input. On the right side an optical splitter (denoted by S) was connected between the amplifier's output and receiver. In this case, when launch power was -1.63~dB (not shown in the table),
the power before amplification is -3.97~dB. In this case, estimated BER at the receiver was $9.35  \times 10^{-15}$. This 
BER value is good compared to the tolerable BER value of $10^{-12}$. However,
when the launch power was reduced to -22.2~dB for the same setup (row 5), the power before amplification was observed as -26.62~dB. The corresponding estimated BER was $3.57  \times 10^{-3}$ and this does not meet the tolerable BER. Thus, it can be observed that when the power level before amplification is below a certain threshold level, signal at the receiver does not meet the tolerable BER. It can also be observed that launch power also has a small but significant role in the network performance.

To illustrate this, let us consider row 16. It has an optical splitter followed by a wavelength multiplexer (denoted by SM) between the transmitter and the amplifier's input. On the other side between the amplifier's output and the receiver, it had a wavelength demultiplexer (denoted by M). In this scenario, the launch power was -19.28~dBm. The observed power level before amplification is -26.64~dBm. This is less than the power level observed with row 5. However, the estimated BER is $4.89 \times 10^{-17}$. This network is able to achieve a good BER.

It can be seen that when the power
before amplification is more than -26.38~dBm, the tolerable BER is
achieved in all scenarios except row 16. Scenario 11 was conducted to confirm the impact of noise
figure. In this scenario, the received power level was
-12.54~dBm. This is less than -12.25~dBm power level expected at the
receiver. It can be seen that in this case, tolerable BER is not
achieved. Thresholds for power level before amplification and 
for received power level were able to handle almost all scenarios but for 
an outlier.

\begin{table}
\centering
{\renewcommand{\arraystretch}{1.1}
\begin{tabular} {|l|l|l|r|r|}
\hline
S.No & LBB & RBB	 & Power (in dBm) &BER
\tabularnewline \hline %3
1& M	& - &-20.38	 &3.01E-15
\tabularnewline \hline %7
2& MM	& - &-12.37	 &4.76E-38
\tabularnewline \hline %19
3 & MM	& - &-24.24	 &8.47E-21
\tabularnewline \hline %6
4& MM	& S &-3.97	 &9.35E-15
\tabularnewline \hline %20
5& MM	& S &-26.62	 &3.57E-03
\tabularnewline \hline %23
6& MS	& - &-29.99	 &9.13E-08
\tabularnewline \hline %2
7& MS	& - & -21.53	 &2.38E-74
\tabularnewline \hline %4
8& MS	& M &-8.53	 &1.51E-14
\tabularnewline \hline %5
9& MS	& S &-6.58	 &1.44E-22
\tabularnewline \hline %21
10& MS	& S &-29.93	 &6.47E-03
\tabularnewline \hline %22
11& MS	& S &-29.99	 &1.67E-04
\tabularnewline \hline %1
12& S & -	 &-18.66	 &2.11E-17
\tabularnewline \hline %10
13& SM	& M &-7.23	 &1.05E-133
\tabularnewline \hline %15
14& SM	& M &-23.94	 &4.41E-22
\tabularnewline \hline %16
15& SM	& M &-24.05	 &8.50E-30
\tabularnewline \hline %1
16& SM	& M &-26.64	 &4.89E-17
\tabularnewline \hline %11*
17*& SM	& S &-11.24	 &9.38E-07
\tabularnewline \hline %12
18& SM	& S &-11.25	 &4.42E-35
\tabularnewline \hline %13
19& SM	& S &-21.53	 &2.63E-29
\tabularnewline \hline %14
20& SM	& S &-26.38	 &6.18E-26
\tabularnewline \hline %8
21& SS	& - &-12.13	 &2.94E-59
\tabularnewline \hline %18
22& SS	& - &-31.76	 &3.48E-10
\tabularnewline \hline %9
23& SS	& M &-10.26	 &5.24E-28
\tabularnewline \hline 
\end{tabular}
}
\caption{Experiment scenarios with amplifier: M denotes wavelength
  multiplexer or demultiplexer and S denotes power combiner or
  splitter. The sequence of letters indicates an interconnection of
  components in the same order. The power level before amplification
  that correlates well with BER is also shown.} 
\label{tbl:edat}
\end{table}

\section{Conclusions}

This paper presented an approach to empirically decide whether an data
center network switch design can operate within tolerable bit error
rates. These designs are multi-fiber short distance networks that are
not widely reported in literature. This paper provides important
experimental data that is required for evaluation of these
networks. It was observed that the BER decision can be made by
considering the optical power at the receiver for networks with
passive optical components. When an amplifier is added to the network,
the BER decision must additionally consider the optical power before
amplification. This model provides a simple way of using thresholds to
evaluate a large number of candidate architectures. This simple model
also eliminates the need for researchers to perform detailed
simulations or experimentation in a large scale.

%\section{TODO}
%Abstract, Introduction about the Infocom paper and the need for model (use motivation), reorder the experiments.
%References: Infocom paper, light runner.

\bibliographystyle{ieeetr}
\bibliography{ref}

\begin{thebibliography}{1}

\bibitem{infocom}
G.~C. Sankaran and K.~M. Sivalingam, ``{Combinatorial Approach for Network
  Switch Design in Data Center Networks},'' in {\em IEEE INFOCOM}, p.~9999,
  2017.

\bibitem{lion}
S.~Yoo, Y.~Yin, and R.~Proietti, ``{Elastic Optical networking and low-latency
  high-radix optical switches for Future Cloud Computing},'' in {\em
  International Conference on Computing, Networking and Communications (ICNC),
  2013}, pp.~1097--1101, IEEE, 2013.

\bibitem{eodcn}
M.~Fiorani, S.~Aleksic, M.~Casoni, L.~Wosinska, and J.~Chen,
  ``{Energy-Efficient Elastic Optical Interconnect Architecture for Data
  Centers},'' {\em IEEE Communications Letters}, vol.~18, pp.~1531--1534, Sept
  2014.

\bibitem{ogdcn}
G.~C. Sankaran and K.~M. Sivalingam, ``Optical traffic grooming-based data
  center networks: Node architecture and comparison,'' {\em IEEE Journal on
  Selected Areas in Communications}, vol.~34, pp.~1618--1630, May 2016.

\bibitem{solver}
G.~C. Sankaran and K.~M. Sivalingam, ``{Grammar based Combinatorial Solver}.''
  \url{https://sourceforge.net/projects/arch-solver/}, Dec. 2015.

\bibitem{ashwin}
A.~Gumaste and T.~Antony, {\em {DWDM network designs and engineering
  solutions}}.
\newblock Cisco Press, 2003.

\bibitem{lightrunner}
\relax{Fiber Optika Technologies}, ``{Light Runner - Fiber Optic Kit}.''
  \url{http://www.fiberoptika.com/fiber-optics-kit.php}, Jan. 2011.

\bibitem{R}
{R Core Team}, {\em R: A Language and Environment for Statistical Computing}.
\newblock R Foundation for Statistical Computing, Vienna, Austria, 2014.

\end{thebibliography}

\end{document}